\documentclass[letterpaper, 10 pt, conference]{ieeeconf}  % Comment this line out
\IEEEoverridecommandlockouts                              % This command is only
\usepackage{dsfont}
\usepackage[utf8]{inputenc}
\usepackage[T1]{fontenc}
\usepackage{graphicx}
\usepackage{subfigure}
\usepackage{dcolumn}% Align table columns on decimal point
\usepackage{bm}% bold math
\usepackage{amsmath}
\usepackage{comment} 
\usepackage{amssymb}
\usepackage{mathrsfs} %jolie lettres
\usepackage[normalem]{ulem}
\usepackage{url}
\usepackage{hyperref}
\usepackage{amsmath}
\usepackage{amssymb}
\usepackage{bbold}
\usepackage[dvipsnames]{xcolor}
\usepackage{xspace}
\usepackage{multirow}

\usepackage{algorithmic}
\usepackage{algorithm2e}

\usepackage[english]{babel} 
\usepackage{caption}
\renewcommand{\thetable}{\arabic{table}}
\AtBeginDocument{%
  
}
\title{\LARGE \bf
Embedding Power Flow into Machine Learning\\ for Parameter and State Estimation
}

\author{Laurent Pagnier and Michael Chertkov% <-this % stops a space
\thanks{This work was supported by M. Chertkov's seed funding at UArizona.}% <-this % stops a space
\thanks{L. Pagnier is with Applied Mathematics at the University of Arizona, Tucson, AZ 85721, USA {\tt\small laurentpagnier@math.arizona.edu}
}
\thanks{M. Chertkov is the Chair of the Program in Applied Mathematics at the University of Arizona, Tucson, AZ 85721, USA and he is also an adjunct professor at the Skolkovo Institute of Science and Technology, Moscow, 121205, Russia {\tt\small chertkov@arizona.edu}}
}

\begin{document}

\maketitle
\thispagestyle{plain}
\pagestyle{plain}

%%%%%%%%%%%%%%%%%%%%%%%%%%%%%%%%%%%%%%%%%%%%%%%%%%%%%%%%%%%%%%%%%%%%%%%%%%%%%%%%
\begin{abstract}
Modern state and parameter estimations in power systems consist of two stages: the outer problem of minimizing the mismatch between network observation and prediction over the network parameters, and the inner problem of predicting the system state for given values of the parameters. The standard solution of the combined problem is iterative: (a) set the parameters, e.g. to priors on the power line characteristics, (b) map input observation to prediction of the output, (c) compute the mismatch between predicted and observed output, (d) make a gradient descent step in the space of parameters to minimize the mismatch, and loop back to (a). We show how modern Machine Learning (ML), and specifically training guided by automatic differentiation, allows to resolve the iterative loop more efficiently. Moreover, we extend the scheme to the case of incomplete observations, where Phasor Measurement Units (reporting real and reactive powers, voltage and phase) are available only at the generators (PV buses), while loads (PQ buses) report (via SCADA controls) only active and reactive powers. Considering it from the implementation perspective, our methodology of resolving the parameter and state estimation problem can be viewed as embedding of the Power Flow (PF) solver into the training loop of the Machine Learning framework (PyTorch, in this study). We argue that this embedding  can help to resolve high-level optimization problems in power system operations and planning.
\end{abstract}

\section{Introduction}
{\bf Parameter Estimation} (PE) and {\bf State Estimation} (SE) are at the core of the Power Systems (PS) operations,  control and optimization \cite{Schweppe1970,2009GCC,2014WWS} at all levels, and specifically at the transmission level this manuscript is focusing on. PE and SE are built in many {\bf Energy Management Systems} (EMS) \cite{abur2004power} 
tasks, such as situational awareness for efficient, comfortable, profitable and secure use, maintenance of equipment, demand management and prediction and short- and long- term planning. 
Modern EMS, implemented in the form of the so-called {\bf Wide Area Measurement Systems} (WAMS) envisioned in 1990s \cite{1985Thorp} and deployed massively in USA \cite{RecoveryAct} and around the globe within the last decade, are cloud based with links to locally hosted {\bf Supervisory Control and Data Acquisition} (SCADA) systems, traditionally (since 1970s) used to control site/node local consumption (or generation) of active and reactive powers automatically, and the {\bf Phasor Measurement Units} (PMUs) \cite{PMU,2010DeLaRee},  synchronized to each other via GPS, positioned at the grid's central locations (e.g. major generators), and reporting local measurements, i.e. (absolute value of) voltage, phase (against a reference point, so called slack bus) and active and reactive injected powers. Integration of the WAMS-PMU technology into the EMS has already generated  many new applications, e.g. in monitoring, state estimation, fault localization and protective relaying, islanding detection (see most recent review \cite{2019Usman} for extended discussions and references).

PS-SE complements analysis of the {\bf Power Flow} (PF) equations, see e.g. Eqs.~(\ref{eq:PF}), by accounting for measurement errors and also taking into account limited observability, i.e. availability of PMU and/or SCADA measurements not at all locations. PS-PE extends PS-PE accounting for parametric uncertainties within PS, e.g. addittance matrix.  Following modern literature on the subject, see e.g. \cite{2013Giannakis} and references there in, SE, PE or the combination of the two, are stated as a data fitting (signal processing) optimization, with many novel solutions reported in recent years \cite{2011Vanfretti,2014Ghiocel,2014ZhuGiannakis,2016Madani}, in particular these taking advantage of the emerging {\bf Machine Learning} (ML) methodologies \cite{2014Barbeiro,2018Mestav,2019WentingLi,zamzam2019datadriven,2019Zhang,2020Wang,pagnier2021physics,2019Bolz,2019Donon,owerko2019optimal,liao2021review,2019Kim,20Liao,Chen_2020} (some of these in the context of a related problem of fault detection).

This manuscript continues the thread, started in our recent paper \cite{pagnier2021physics} and in a companion submission to CDC \cite{2021Afonin}, suggesting application of the {\bf Physics-Informed} ML (PIML) to SE and PE in PS models. Essence of PIML is in resolving the problem as a regression based on the PF equations in their static or dynamic (then working with the so-called swing equations) \cite{machowski1997power,huang2009application,chiang2011direct,zhou2011calibration,guo2014adaptive,zhou2015dynamic,chen2016measurement,chavan2017identification,wang2017pmu,2018LearningPowerSystemDynamics,ostrometzky2020physicsinformed}.
We focus here on the static PF setting, therefore assuming access to a static (or quasi-static) data consistent with solutions of the PF Eqs.~(\ref{eq:PF}). 

In our prior work on the subject \cite{pagnier2021physics} we resolve the SE and ME by training a ML scheme to reproduce the explicit {\bf Inverse Map} (function) from amplitudes and phases of the voltage potentials to complex powers at the nodes of the PS (that is utilizing the map formally defined below in Section \ref{sec:PF} as $\bm \Pi^{-1}$). This specific choice of the input and output in the supervised learning setting of \cite{pagnier2021physics} was made to turn the PF equations into an explicitly executable map (i.e. map not requiring to solve an nonlinear multi-parametric algebraic equation).  Even thought the approach is appealing technically it is also handicapped in terms of the PS applications because
\begin{itemize}
    \item sampling of the training data is distorted in comparison with the practical input-output setting (see Eq.~(\ref{eq:PF-IO}) below);
    \item the approach does not address the fact that solving the PF is the process embedded directly in all the higher level optimization and control tasks,  such as finding optimal generator dispatch and other \cite{2009GCC,2014WWS}. 
\end{itemize}
Desire to correct for the problem motivates the alternative approach this manuscript suggests: embedding the standard way of solving the PF Eqs.~(\ref{eq:PF}),  via the {\bf Newton-Raphson} (NR) scheme \cite{arthur2000power,grainger1994power}, into a {\bf PIML algorithm}. 

We show in this manuscript that: 
\begin{itemize}
    \item The NR scheme with a fixed number of iterations can be built in the ML algorithm explicitly thus allowing to take advantage of the automatic differentiation for efficient training (back propagation). We coin it the {\bf Newton-Raphson informed Machine Learning} (NR-ML) algorithm.
    \item Only a few iterations of the NR scheme, hard-coded within NR-ML, are sufficient to get high-quality SE and PE reconstruction.
    \item SE is achieved relatively fast, 
    while a longer training is required to achieve acceptable quality of PE. 
    \item The Power System Informed approach taken in the manuscript allows to generalize/extrapolate, i.e. achieve a very good quality of the SE reconstruction even when tested in the regimes (on samples) which are sufficiently far from these used in training. 
\end{itemize}

The material in the remainder of the manuscript is organized as follows. We set nomenclature (e.g. for the NR solution of the PF problem) in Section \ref{sec:PF}. Specifics of the NR-ML algorithm, e.g. Input-Output data, PE prescribed Loss Function for training and SE validation test, are described in Section \ref{sec:ML}. Our training and validation experiments with NR-ML on an exemplary 118-node IEEE model 
are described in Section \ref{sec:experiments}. Section \ref{sec:conclusions} is reserved for conclusions and discussions of the path forward.

\section{Solving the Power Flow Equations}\label{sec:PF}

We consider a transmission level power grid over the grid-graph, ${\cal G}=({\cal V},{\cal E})$, where ${\cal V}$ and ${\cal E}$ are the set on nodes (generators or loads) and set of lines, respectively. The Power Flow (PF) equations, governing steady redistribution of power over the system, are stated in terms of the complex injected powers, $\forall i\in{\cal V}:\ {S}_i\equiv p_i+{\bm i} q_i$, and in terms of the complex potentials, $\forall i\in{\cal V}:\ {V}_i\equiv v_{i}\exp({\bm i}\theta_i)$, where ${\bm i}^2=-1$; $v_i$ and  $\theta_i$ denote voltage (magnitude) and phase of the (voltage) potential at the node $i$: $\forall i\in{\cal V}$,
\begin{gather}\label{eq:PF}
    \begin{array}{l}p_i=\mathcal{P}_i(\bm v, \bm \theta)\equiv\underset{j\in \mathcal{V}}{\sum}v_iv_j\big(\,g_{ij}\cos\theta_{ij}+b_{ij}\sin\theta_{ij}\,\big)\vspace{2pt},\\ 
    q_i=\mathcal{Q}_i(\bm v, \bm \theta)\equiv\underset{j\in\mathcal{V}}{\sum}v_iv_j\big(\,g_{ij}\sin\theta_{ij}-b_{ij}\cos\theta_{ij}\,\big),
\end{array}
\end{gather}
where $\theta_{ij} \equiv \theta_{i} - \theta_{j}$ and  $\bm Y=\bm g +\bm i \bm b$ is the admittance matrix. 
The equation~(\ref{eq:PF}) can be viewed as the map from the complex powers (input) to the complex potentials, ${\bm \Pi}: {\bm S}\equiv ({S}_i|i\in{\cal V})\mapsto {\bm V}\equiv ({V}_i|i\in{\cal V})$, which is non-injective and implicit and may allow no solution or have multiple solutions. It can also be interpreted in terms of the inverse map, ${\bm \Pi}^{-1}: {\bm V}\mapsto {\bm S}$, which is the one utilized in the ML schemes of \cite{pagnier2021physics}. 
(It is injective and explicit function mapping the complex potentials (input) to complex powers (outputs).) In the (transmission system) practical setting inputs and outputs in Eqs.~(\ref{eq:PF}) are mixed, depending on if a node corresponds to a generator, $i\in{\cal V}^{(g)}$, or a load, $i\in{\cal V}^{(l)}$, where ${\cal V}^*={\cal V}\setminus 0={\cal V}^{(g)}\cup {\cal V}^{(l)}$ and ${\cal V}^{(g)}\cap {\cal V}^{(l)}=\emptyset$:
\begin{gather}\label{eq:PF-IO}
\begin{array}{rrr}
\text{Input}:& (p_i,v_i)\text{ if }i\in {\cal V}^{(g)};& (p_i,q_i)\text{ if }i\in {\cal V}^{(l)};\\
\text{Output}:& (q_i,\theta_i)\text{ if }i\in {\cal V}^{(g)};& (v_i,\theta_i)\text{ if }i\in {\cal V}^{(l)},
\end{array}
\end{gather}
and $0$ is a special bus -- the so-called slack bus (usually associated with the largest generator in the system) where phase and voltage are fixed according to, $\theta_0=0$ and $v_0=1$. 
In this manuscript we work with the (practical) PF setting, described by Eqs.~(\ref{eq:PF},\ref{eq:PF-IO}) and also supplemented by the slack node conditions, which is coined the {\bf Mixed-PV-PQ} map. (This name reflects on the fact that, according to Eq.~(\ref{eq:PF-IO}) the map has a mixed input -- $(p_i,v_i|i\in{\cal V}^{(g)})$ for the generators and $(p_i,v_i|i\in{\cal V}^{(g)})$ for the loads). Notice, that the Mixed-PV-PQ may have unique, no or multiple solutions, like in the inverse map $\bm \Pi^{-1}$ setting described above. 

Newton-Raphson (NR) scheme is normally used to resolve the PF equations in the Mixed-PV-PQ setting. The NR scheme works as follows. The admittance matrix, ${\bm Y}$,
vector of active power injections/consumptions at the loads and generators, ${\bm p}\equiv(p_i|i\in{\cal V}^*)$, vector of the reactive demands at the loads, ${\bm q}^{(l)}\equiv(p_i|i\in{\cal V}^{(l)})$, and vector of voltage magnitudes at the generators, $\bm v^{(g)}\equiv(v_i|i\in{\cal V}^{(g)})$, are fixed. We introduce the vector of voltage magnitudes at the loads, ${\bm v}^{(l)}\equiv(v_i|i\in{\cal V}^{(l)})$, the vector of phases, ${\bm \theta}\equiv(\theta_i|i\in{\cal V}^*)$,  the vectors of the active and reactive power mismatches, $\Delta \bm p\equiv(\mathcal{P}_i(\bm v, \bm \theta)-p_i|i\in {\cal V}^*)$ and $\Delta \bm q\equiv(\mathcal{Q}_i(\bm v, \bm \theta)-q_i|i\in {\cal V}^{(l)})$, respectively, which are updated at each step of the NR iterations as follows
\begin{equation}\label{eq:NR}
\left[\begin{array}{c}
\!\!\Delta \bm \theta\!\!\\
\!\!\Delta \bm v\!\!\\
\end{array}
\right]=
\bm J\big(\bm v, \bm \theta\,; \bm Y\big)^{-1}\cdot
\left[\begin{array}{c}
\!\!\Delta \bm p\!\!\\
\!\!\Delta \bm q\!\!\\
\end{array}
\right]\,,
\end{equation}
where, $\Delta \bm v\equiv (\Delta v_i|i\in{\cal V}^{(l)})$ and $\Delta \bm \theta\equiv (\Delta \theta_i|i\in{\cal V}^*)$, are increments of the voltage and phase vectors, respectively, acquired (during the elementary NS step). $\bm J$ in Eq.~(\ref{eq:NR}) is the Jacobian matrix of the PF equations, which is convenient to decompose into sub-matrices according to 
\begin{equation}
\bm J = \left[\begin{array}{cc}
\!\!\bm J^{\mathcal{P}\theta}&\!\!\bm J^{\mathcal{P}v}\!\!\\
\!\!\bm J^{\mathcal{Q}\theta}&\!\!\bm J^{\mathcal{Q}v}\!\!
\end{array}\right]\,,
\end{equation}
where $\bm J^{\mathcal{P}\theta}\equiv (\partial \mathcal{P}_i/\partial \theta_j|i,j\in {\cal V}^*)$, $\bm J^{\mathcal{P}v}\equiv (\partial \mathcal{P}_i/\partial v_j|i\in {\cal V}^*,\ j\in{\cal V}^{(l)})$,
$\bm J^{\mathcal{Q}\theta}\equiv (\partial \mathcal{Q}_i/\partial \theta_j|i\in{\cal V}^{(l)},j\in {\cal V}^*)$, $\bm J^{\mathcal{Q}v}\equiv (\partial \mathcal{Q}_i/\partial v_j|i\in{\cal V}^{(l)},\ j\in{\cal V}^{(l)})$.  
Eq.~(\ref{eq:NR}) should also be complemented by (direct) computation of the vector of the reactive powers injected (or consumed) at the generators, $\bm q^{(g)}=(\mathcal{Q}_i(\bm v,\bm \theta)|i\in{\cal V}^{(g)})$. See  Algorithm~\ref{alg:NR} for the pseudo-code of the NR algorithm repeated $n$ times within the Mixed-PV-PQ setting. The algorithm is initialized with the ``flat start'': $v_i^{(0)}=v_i^{(g)}\,, \forall i \in \mathcal{V}^{(g)}$, $v_i^{(0)}=1\,, \forall i \in \mathcal{V}^{(l)}$ and $\theta_i^{(0)}=0\,, \forall i\in \mathcal{V}$,  where $v_i^{(g)},\ \forall i\in{\cal V}^{(g)}$, are voltage set-points at the generators (in the dimensionless units).
\begin{algorithm}[h]
\SetAlgoLined
\SetKwInOut{Input}{Input}
\SetKwInOut{Output}{Output}
\Input{$\bm v^{(g)}$, $\bm p$, $\bm q^{(l)}$, $\bm Y$}
\Output{$\bm \theta^{(n)}$, $\bm v^{(n;l)}$, $\bm q^{(n;g)}$}
initialize $\bm v^{(0;l)}=(1|i\in{\cal V}^{(g)})$, $\bm \theta^{(0)}=(0|i\in{\cal V})$\\
 \For{$k=1$ \KwTo n}{
 compute $\Delta \bm p= \bm{\mathcal{P}}(\bm v, \bm \theta)-\bm p$,\\
 \phantom{compute }$\Delta \bm q= \bm{\mathcal{Q}}(\bm v, \bm \theta)-\bm q$, \vspace{3pt}\\
solve $\bm J\big(\bm v^{(k-1)}, \bm \theta^{(k-1)};\,\bm Y\big)\cdot \left[\begin{array}{c}\!\!\!\Delta \bm \theta\!\\ \!\!\Delta \bm v\end{array}\!\!\right]=\left[\begin{array}{c}\!\!\Delta \bm p\!\!\\\!\!\Delta \bm q\!\!\end{array}\right]$,\vspace{3pt}\\
update $\bm \theta^{(k)}= \bm \theta^{(k-1)}+\Delta \bm \theta$,\\
    \phantom{update }$\bm v^{(k;l)}= \bm v^{(k-1;l)}+\Delta \bm v$,\\

 }
     
 compute $\bm q^{(n;g)}= (\mathcal{Q}_i(\bm v^{(n)},\bm \theta^{(n)})|i\in{\cal V}^{(g)})\,.$ \vspace{5pt}

 \caption{$n$-step Newton-Raphson scheme of the Mixed-PV-PQ setting. $\text{NR}^{(n;g)}_{\bm Y}\big(\bm v^{(g)},\bm p,\bm q^{(l)}\big)=\big(\bm v^{(n;g)}; \bm \theta^{(n;g)}\big)$,
is the algorithm's output vector at the generator also indicating dependence on the input parameters.
}\label{alg:NR} 
\end{algorithm}

\section{Machine Learning Framework}\label{sec:ML}
We assume that the PMUs are installed at generators, therefore providing observations of active and reactive powers, voltages and phases at all the generators and that power injections (active and reactive) are observed at the load nodes (equipped with SCADA sensors):
\begin{equation*}
\text{measurements}:\left\{\begin{array}{l}v_i, \theta_i, p_i, q_i, \hspace{10pt}\forall i \in \mathcal{V}^{(g)}\,,\\
p_i, q_i, \hspace{35pt}\forall i \in \mathcal{V}^{(l)}\,.
\end{array}\right.
\end{equation*}

Therefore we generate $N$ observation sets, each labeled by $\alpha=1,\cdots,N$, and minimize the following loss function over the vector of parameters (the admittance matrix $\bm Y$):
\begin{align}\label{eq:Loss}
&\min\limits_{\bm Y}\mathcal{L}(\bm Y),\quad \mathcal{L}(\bm Y)\equiv\\
\nonumber &\quad \frac{1}{N|\mathcal{V}^{(g)}|}\sum_{\alpha=1}^{N}
\left\|
\left[\!\!\!\begin{array}{c}
 \vphantom{v^{(g)}_\alpha}\multirow{2}{*}{$\text{NR}^{(n;g)}_{\bm Y}\Big(\bm v^{(g)}_\alpha,\bm p_\alpha,\bm q^{(l)}_\alpha\Big)$}\\
 \vphantom{\theta^{(g)}_\alpha}\\
\bm{\mathcal{P}}^{(g)}\big(\bm v_\alpha, \bm\theta_\alpha\big)
\end{array}\!\!\!\right]
-\left[\!\!\begin{array}{l}
\bm v^{(g)}_\alpha\\
\bm \theta^{(g)}_\alpha\\
\bm p_\alpha^{(g)}
\end{array}\!\!\right]
\right\|^2\,.
\end{align}
The minimization is performed with Adam optimizer \cite{Adam} (an algorithm for first-order gradient-based optimization of stochastic objective functions based on adaptive estimates of lower-order moments) in PyTorch \cite{PyTorch}, utilizing for faster convergence tensor representation, automatic differentiation over the grid parameters, $\bm Y$, and alternating between batches of samples at each backward propagation epoch. See Appendix~\ref{sec:batch} for implementation details.

\section{Experiments}\label{sec:experiments}
We experiment with the IEEE-118 model. We follow the data generation process described in \cite{pagnier2021physics} and prepare 
five different data sets. Each data set consists of 2000 samples. The data sets are sorted from case \#1 to case \#5 in terms of their perceived mapping complexities. As observed in \cite{pagnier2021physics}, the Physics-Informed ML scheme requires much smaller training sets in comparison with their physics-agnostic counterparts. This observation extends to the present work. Here we assign one out of fifty samples to the training set, the remaining samples are used for validation.
\begin{figure}[h!]
\includegraphics[width=\columnwidth]{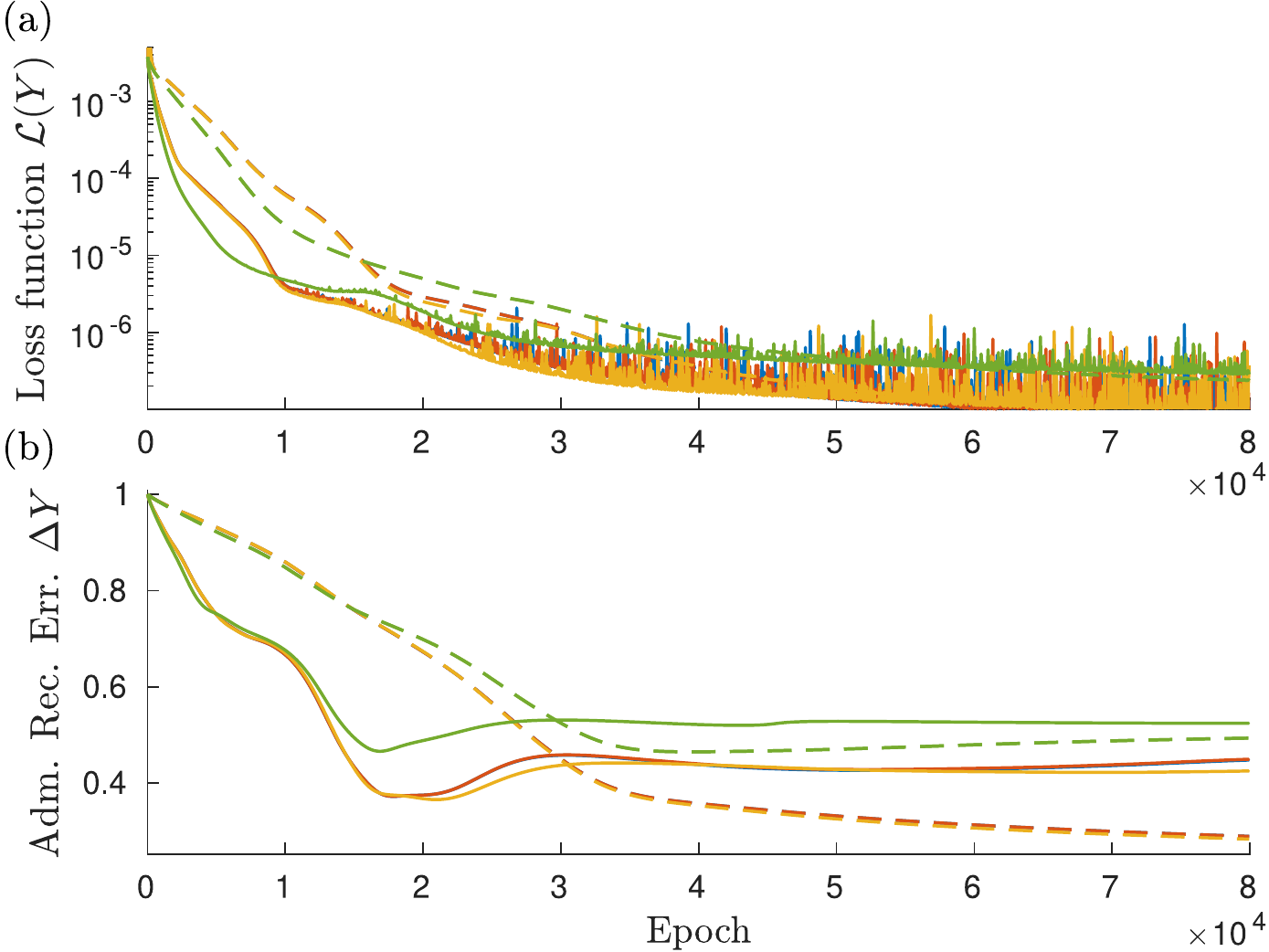}
\caption{Dependence of the loss function, $\mathcal{L}(\bm Y)$, and of the reconstruction error, $\Delta \bm Y$, on the number epochs (training progression) for case \#1 with $n=1$ (green), $n=2$ (orange), $n=3$ (red), $n=4$ (blue) for $l_r=5\cdot10^{-4}$ (solid) and  $l_r=1\cdot10^{-4}$ (dashed). Curves for $n=3$ and $4$ are almost indistinguishable.}\label{fig:train}
\end{figure}

Dependence of the Loss Function (LF), $\mathcal{L}(\bm Y)$, on the number of epochs is selected as an indicator of the quality of training.  Fig~\ref{fig:train} (a) shows that the LF decreases monotonically and a rather small values of  are reached relatively fast.  Moreover,  the monotonic decrease of the LF with the training time (number of epochs) is observed at any number of NR steps (even at $n=1$) hard-coded into the training process and also for different values of the learning rate $l_r$. We notice that decay of the LF saturates to a small value in the process of training relatively fast, as seen in Fig~\ref{fig:train} (a). This observation is also confirmed in the validation test (see Table \ref{tab:n_iter}),  therefore reporting State Estimation of a good quality.  To test the Parameter Estimation, which measures quality of reconstructing the admittances, we analyze the Admittance Reconstruction Error (ARE)
\begin{equation}
\Delta \bm Y = \frac{1}{\mathcal{N}}\big\|\bm Y- \bm Y_{\rm ref}\big\|_{F}=\frac{1}{\mathcal{N}}\Big(\sum_{i,j\in \mathcal{V}} \big|y_{ij}-y_{ij}^{\rm ref}\big|^2\Big)^{\frac{1}{2}}\,,
\end{equation}
where $\mathcal{N}$ is the reconstruction error obtained with the initial values of the grid parameters and $\bm Y_{\rm ref}$ is the actual admittance matrix. 
Fig~\ref{fig:train} (b), showing dependence of the ARE on the number of the training epochs, suggests that reconstruction of the State Estimation (SE), controlled by the Loss Function,  is observed relatively early in the training process when the Parameter Estimation (PE),  controlled by the ARE,  is still large. ARE decreases  with increase in the number of epochs leading, eventually,  to the PE of a sufficiently high quality. We also observe in Fig~\ref{fig:train} (b) a consistent decrease with the number of the NR iterations, $n$, and also a stronger dependence on $n$ than one seen in the LF.  
At $n=1$, the ARE plateaus at a much higher value than for the schemes with $n>1$.

We have also experimented with the learning rate $l_r$ and observed that it has a strong influence on both SE and PE. We have reached the conclusion,  that even though increasing learning rate may speed up decrease of the ARE at the early stages of training, the aggressive strategy is sub-optimal overall as it does not result in the LF reaching its minimal value. Moreover, when we increase the number of epochs (in the discussed case of the high learning rate) $\mathcal{L}$ continues to decrease while $\Delta \bm Y$ starts to increase.  We relate this trouble to the overfitting phenomenon,  and plan to continue the experimental work in the future in order to find an empirically satisfactory stopping criterion. 

Following protocol which is standard in ML we conduct extensive validation tests on the samples not used for training. Such tests are important to quality of the SE in terms of the ability to generalize, i.e. extrapolate into regimes not seen in training. Specifically, we introduce the following Validation Error, testing the quality of prediction at the loads: 
\begin{align}
E&\equiv \frac{1}{N|\mathcal{V}|}\sum_{\alpha=1}^{N}\Bigg(\left\|\text{NR}^{(n;g)}_{\bm Y}\Big(\bm v_\alpha^{(g)},\bm p_\alpha,\bm q_\alpha^{(l)}\Big) - \left[\!\!\begin{array}{c}\bm \theta_\alpha^{(g)}\\\bm q_\alpha^{(g)}\end{array}\!\!\!\right]\right\|^2\nonumber\\
&+\left\|\text{NR}^{(n;l)}_{\bm Y}\Big(\bm v_\alpha^{(g)},\bm p_\alpha,\bm q_\alpha^{(l)}\Big)  - \left[\!\!\begin{array}{c}\bm \theta_\alpha^{(l)}\\\bm v_\alpha^{(l)}\end{array}\!\!\!\right]\right\|^2\Bigg)\,.\label{eq:valid}
\end{align}
Dependence of the Training Error (Loss Function), Validation Error, Reconstruction Error and   Duration (normalized by the training time for $n=1$) on the number of NR iterations, $n$, are reported in Table~\ref{tab:n_iter}. 
The results shown in the Table were collected for the case of the  learning rate, $l_r=10^{-4}$,  and the number of epochs, $N_{\rm e} = 8\cdot 10^{4}$.
We observe that the training time increases roughly linearly with the number of NR iterations. On the other hand, we also report that the accuracy of the results stops to improve after $n=3$.

\begin{figure}[h!]
\includegraphics[width=\columnwidth]{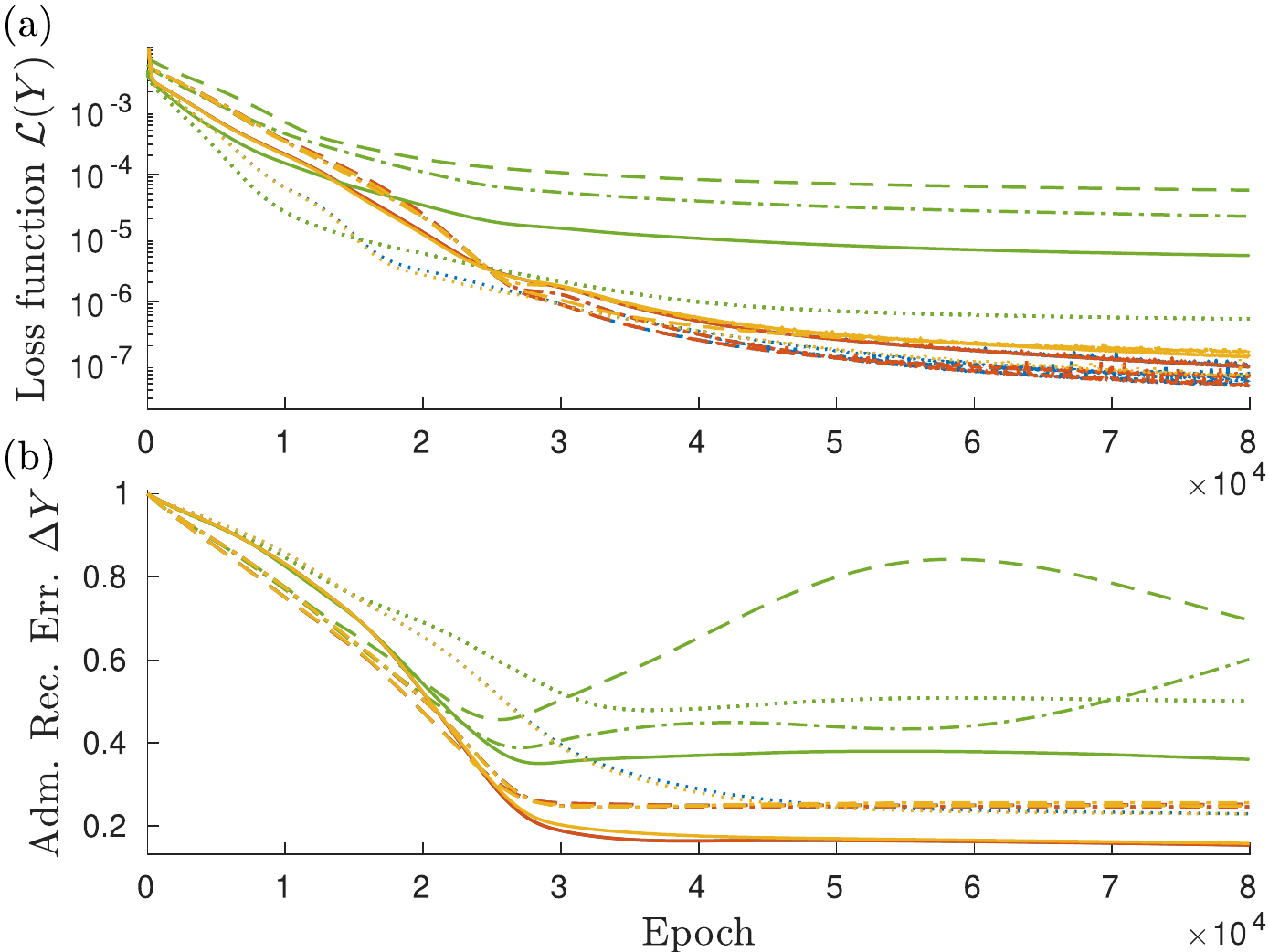}
\caption{Evolution of the Loss Function, $\mathcal{L}(\bm Y)$, and of the Reconstruction Error, $\Delta \bm Y$, during training for $n=1$ (green), $n=2$ (orange), $n=3$ (red), $n=4$ (blue) when trained over case \#2 (dotted), case \#3 (solid), case \#4 (dashed) and case \#5 (dash-dot) with a learning rate $l_r=1\cdot10^{-4}$.}\label{fig:diff_set}
\end{figure}

\begin{figure}[h!]
\includegraphics[width=\columnwidth]{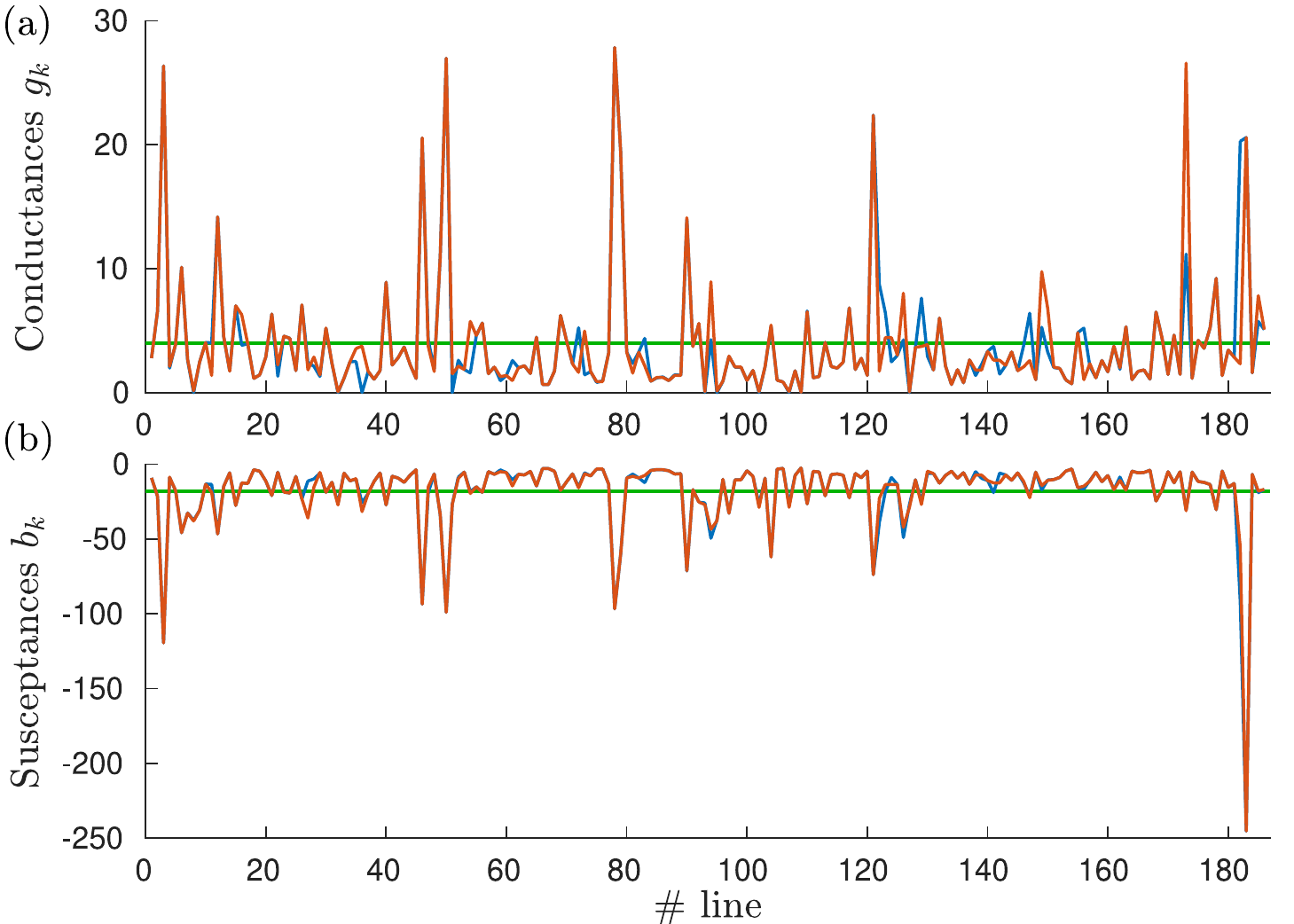}
\caption{Line conductances and susceptances are displayed in panels (a) and (b) respectively. Their initial (pre-training) values are displayed in green, while their trained values and their reference counterparts are displayed in red and blue respectively. These results are obtained for case \#3 with $n=3$, $l_r=10^{-4}$ and $N_{\rm e} = 8\cdot 10^{4}$.}\label{fig:grid_param}
\end{figure}

\begin{table}[h!]
\caption{Validations test: comparison of different $n$ of iterations in N-R scheme.}\label{tab:n_iter}
\center
\begin{tabular}{ccccc}
\hline
NR iter. & Train.  & Valid. & Adm. Rec. & Duration\\
$n$ & Err. $\mathcal{L}$ & Err. $E$ & Err. $\Delta Y$ & \\
\hline
1& 6.79E-7 & 7.68E-3 & 0.493 & 1.00\\
2& 5.33E-7 & 1.42E-3 & 0.284 & 2.36\\
3& 5.35E-7 & 1.16E-3 & 0.290 & 3.52\\
4& 5.34E-7 & 1.15E-3 & 0.290 & 4.24\\
6& 5.36E-7 & 1.15E-3 & 0.290 & 6.45\\
8& 5.35E-7 & 1.15E-3 & 0.290 & 8.30\\
\hline
\end{tabular}
\end{table}
In Fig.~\ref{fig:diff_set}, we compare how our method perform on different data sets. Paradoxically, case \#3 gives a slightly better reconstruction of $\bm Y$ while having a slightly higher loss function values. Finally, Fig.~\ref{fig:grid_param} shows a comparison between the estimated and real line admittances. The main message here is that the agreement is almost perfect. We also note that for these few lines where the reconstruction of admittances is not perfect,  the reconstructed admittances are reported within their physically sensible ranges (positive for conductances and negative for susceptances). We attribute this success of the Parameter Estimation  to the exponential parametrization of the susceptances. (See Appendix~\ref{sec:param} for details.)

\section{Conclusions and Path Forward}\label{sec:conclusions}

In this manuscript we showed how the Newton-Raphson scheme of resolving the Power Flow equation, which is a key tool of standard power system analysis, can be used within ML technique producing efficiently State Estimation and, in parallel, efficient (and interpretable) Parameter Estimation based on the PMU and SCADA data sources. Advantage of the approach, in comparison with other alternatives recently discussed in the literature (in particular by the authors of this manuscript) is that the Input-Output, Mixed PV-PQ framework is much closer to the practical  engineering reality of the power system operations, control and planning. We therefore believe that the methodology suggested will become an important building block of future development of a number data driven optimization scheme, such as data aware (driven) generation and demand respond dispatch, unit commitment and multi-scenario planning. 

Most significant technical result of our approach consists in building in the Newton-Raphson (NR) scheme within a modern ML framework, specifically PyTorch, which has allowed us to take advantage of the automatic differentiation and related technical tools developed by the ML community in recent years. Quite remarkably our experiments show that hard-coding into the training procedure only 2-3 iterations of the NR is sufficient for getting high quality State and Parameter Estimations while  keeping computational time and memory requirements relatively light.

We also encountered an implementation problem which requires attention of the ML community developing PyTorch (or other similar products). Our attempt to scale up computations to larger grids run into memory problem, because PyTorch does not yet support sparse (linear algebra) computations,  specifically it is lacking ability to solve efficiently large but sparse system of equations (invert large sparse matrices). 
Building ML computational framework which allows compression of the sparse data structures, such as these associated with the tensor blocks our algorithm is built on (see Appendix~\ref{sec:batch}), is important for making the Power System Informed (and more generally Physics Informed) ML schemes practical. 

\appendix

\subsection{Batch version of the Mixed-PV-PQ Newton-Raphson  Scheme}\label{sec:batch}

ML schemes are usually trained over (mini-)batches. Batch approximation of the gradient reduces oscillation and noise during the gradient descent and therefore provides a more stable implementation. Most algorithms of the modern ML, e.g. implemented in PyTorch, have been optimized to handle tensor inputs because tensor operations are typically much more efficient then nested loops.

In this Appendix we give technical details, which are critical for efficient implementation, on the tensor operations needed to train the Mixed-PV-PQ Newton Raphson scheme. 

Let us define the source incidence matrix 
\begin{equation}
B_{ik}^{\,\rm out} = \left\{\begin{array}{l}
1\,, \text{ if the node $i$ the source of the directed line $k$,}\\
0\,, \text{ otherwise,}
\end{array}\right.
\end{equation}
and the target incidence matrix 
\begin{equation}
B_{ik}^{\,\rm in} = \left\{\begin{array}{l}
1\,, \text{ if bus $i$ is the target of the directed line $k$,}\\
0\,, \text{ otherwise,}
\end{array}\right.
\end{equation}
then the undirected and directed incidence matrices $\bm B^{+}$ and $\bm B^{-}$ read
\begin{align}
\bm B^{+} &= \bm B^{\,\rm in} + \bm B^{\rm out}\,,\\
\bm B^{-} &= \bm B^{\,\rm in} - \bm B^{\rm out}\,.
\end{align}
In the following we are using Einstein summation convention (summation over repetitive indexes). Components of the admittance matrix $\bm Y$ can be written in terms of the line admittances and shunt admittances as follows
\begin{equation}
y_{ij} = B_{ik}^{-}y_k B_{jk}^{-} + \delta_{ij}y_j^{\rm sh}\,,
\end{equation}
where $y_k=g_k +\bm i b_k$ is the admittance of line $k$ and $y_j^{\rm sh}$  is the shunt admittance at bus $j$. Let us introduce the following two useful elementary tensors 
\begin{align}
c_{k \alpha} &= \cos\big(B_{ik}^{-}\theta_{i \alpha}\big)\,,\\
s_{k \alpha} &= \sin\big(B_{ik}^{-}\theta_{i \alpha}\big)\,.
\end{align}
Then active and reactive injections based on $v_{i\alpha}$ and $\theta_{i\alpha}$ are
\begin{align}
\mathcal{P}_{i\alpha} &= v_{i\alpha}\Big[B_{ik}^{\,\rm in}\big(-g_{k}c_{k\alpha}-b_{k}s_{k\alpha}\big)B_{jk}^{\,\rm out}\nonumber\\
& + B_{ik}^{\,\rm out}\big(-g_{k}c_{k\alpha}+b_{k}s_{k\alpha}\big) B_{jk}^{\,\rm in}\nonumber\\
& + \delta_{ij}(B_{jk}^{+}g_k + g_j^{\rm sh})\Big] v_{j\alpha}\,,\\
\mathcal{Q}_{i\alpha} &= v_{i\alpha}\Big[B_{ik}^{\,\rm in}\big(-g_{k}s_{k\alpha}+b_{k}c_{k\alpha}\big)B_{jk}^{\,\rm out}\nonumber\\
& + B_{ik}^{\,\rm out}\big(+g_{k}s_{k\alpha}+b_{k}c_{k\alpha}\big) B_{jk}^{\,\rm in}\nonumber\\
& - \delta_{ij}(B_{jk}^{+}b_k + b_j^{\rm sh})\Big] v_{j\alpha}\,,
\end{align}
where $\delta_{ij}$ is the Kronecker delta. Tensors contributing the Jacobian are 
\begin{align}
J_{ij\alpha}^{\mathcal{P}\theta} & = v_{i\alpha}\Big[B_{ik}^{\,\rm in} \big(-g_{k}s_{k\alpha}+b_{k}c_{k\alpha}\big)B_{jk}^{\,\rm out} \nonumber\\\
&+ B_{ik}^{\,\rm out} \big( g_{k}s_{k\alpha}+b_{k}c_{k\alpha}\big) B_{jk}^{\,\rm in} + \delta_{ij}j_{i\alpha}^{P\theta}\Big] v_{j\alpha}\,,\\
J_{ij\alpha}^{\mathcal{P}v} & = v_{i\alpha}\Big[B_{ik}^{\,\rm in} \big(-g_{k}c_{k\alpha}-b_{k}s_{k\alpha}\big)B_{jk}^{\,\rm out} \nonumber\\\
&+ B_{ik}^{\,\rm out} \big( -g_{k}c_{k\alpha}+b_{k}s_{k\alpha}\big) B_{jk}^{\,\rm in} + \delta_{ij}j_{i\alpha}^{PV}\Big] \,,\\
J_{ij\alpha}^{\mathcal{Q}\theta} & = v_{i\alpha}\Big[B_{ik}^{\,\rm in} \big(g_{k}c_{k\alpha}+b_{k}s_{k\alpha}\big)B_{jk}^{\,\rm out} \nonumber\\\
&+ B_{ik}^{\,\rm out} \big(g_{k}c_{k\alpha}-b_{k}s_{k\alpha}\big) B_{jk}^{\,\rm in} + \delta_{ij}j_{i\alpha}^{Q\theta}\Big] v_{j\alpha}\,,\\
J_{ij\alpha}^{\mathcal{Q}v} & = v_{i\alpha}\Big[B_{ik}^{\,\rm in} \big(-g_{k}s_{k\alpha}+g_{k}c_{k\alpha}\big)B_{jk}^{\,\rm out} \nonumber\\\
&+ B_{ik}^{\,\rm out} \big( g_{k}s_{k\alpha}+b_{k}c_{k\alpha}\big) B_{jk}^{\,\rm in} + \delta_{ij}j_{i\alpha}^{PV}\Big] \,,
\end{align}
with
\begin{align}
J_{i\alpha}^{\mathcal{P}\theta} &= v_{i\alpha}\Big[B_{ik}^{\,\rm in} \big(g_{k}s_{k\alpha}-b_{k}c_{k\alpha}\big)B_{jk}^{\,\rm out} \nonumber\\\
&+ B_{ik}^{\,\rm out} \big(-g_{k}s_{k\alpha}-b_{k}c_{k\alpha}\big) B_{jk}^{\,\rm in}\Big] v_{j\alpha}\,,\\
J_{i\alpha}^{\mathcal{P}v} & = \Big[B_{ik}^{\,\rm in} \big(-g_{k}c_{k\alpha}-b_{k}s_{k\alpha}\big)B_{jk}^{\,\rm out} \nonumber\\\
&+ B_{ik}^{\,\rm out} \big( -g_{k}c_{k\alpha}+b_{k}s_{k\alpha}\big) B_{jk}^{\,\rm in}\Big]v_{j\alpha} \nonumber\\
&+2(B_{ik}^{+}g_k+g_i^{\rm sh})v_{i\alpha}\,,\\
J_{i\alpha}^{\mathcal{Q}\theta} & = v_{i\alpha}\Big[B_{ik}^{\,\rm in} \big(-g_{k}c_{k\alpha}-b_{k}s_{k\alpha}\big)B_{jk}^{\,\rm out} \nonumber\\\
&+ B_{ik}^{\,\rm out} \big(-g_{k}c_{k\alpha}+b_{k}s_{k\alpha}\big) B_{jk}^{\,\rm in}\Big] v_{j\alpha}\,,\\
J_{i\alpha}^{\mathcal{Q}v} & = \Big[B_{ik}^{\,\rm in} \big(-g_{k}s_{k\alpha}+g_{k}c_{k\alpha}\big)B_{jk}^{\,\rm out} \nonumber\\\
&+ B_{ik}^{\,\rm out} \big( g_{k}s_{k\alpha}+b_{k}c_{k\alpha}\big) B_{jk}^{\,\rm in}\Big]v_{j\alpha}\nonumber\\
&- 2(B_{ik}^{+}b_k+b_i^{\rm sh})v_{i\alpha}\,.
\end{align}

\subsection{Parameterization of line admittances}\label{sec:param}

It is highly advisable in any ML approach to adjust different parameters to make their initializations (priors), ranges and expected values comparable. However bare variation in line admittances over the system may spread over more than a decade. To overcome this difficulty we represented in  \cite{pagnier2021physics} line admittances via resistances and reactances
\begin{align}
g_k&=r_k\big/\big(r_k^2+x_k^2\big)\,,\\
b_k&=-x_k\big/\big(r_k^2+x_k^2\big)\,,
\end{align}
taking advantage of the fact that the latter characteristics spread over much narrower range. One wishes reconstructed admittances to be within their physically sensible ranges, this can be achieved by adding the following regularization term to the loss function
\begin{equation}
\mathcal{R}(\bm Y) = \lambda \sum_k \big[\max\big(0,-r_k\big) + \max\big(0,-x_k\big)\big]\,.
\end{equation}
In this manuscript, we have implemented what we believe is a simpler solution, as achieving the same goal however without regularization. We simply use the following ``exponential'' parameterization
\begin{align}
g_k&=\exp(\gamma_k)\,,\\
b_k&=-\exp(\beta_k)\,.
\end{align}

\end{document}